\documentclass[draftclsnofoot, onecolumn]{IEEEtran}

\usepackage[final]{graphicx}
\usepackage{subfigure}
\usepackage{amsmath,amssymb}
\usepackage{eso-pic}
\usepackage{stfloats}
\usepackage{multirow}
\usepackage{footnote}
\usepackage[hyphens]{url}
\usepackage[finalnew]{trackchanges}
\ifCLASSINFOpdf
\else
\fi

\begin{document}
\title{Digital Prototype Filter Alternatives for Processing Frequency-Stacked Mobile Sub-Bands Deploying a Single ADC for Beamforming Satellites}

\author{
	\IEEEauthorblockN{Adem Coskun\IEEEauthorrefmark{1}, Sevket Cetinsel\IEEEauthorrefmark{2}, Izzet Kale\IEEEauthorrefmark{3}, Robert Hughes\IEEEauthorrefmark{2}, Piero Angeletti\IEEEauthorrefmark{1}, and Christoph Ernst\IEEEauthorrefmark{1}}
	\IEEEauthorblockA{\\ \IEEEauthorrefmark{1}European Space Agency, Noordwijk, The Netherlands
		\\\{adem.coskun, piero.angeletti, christoph.ernst\}@esa.int}
	\IEEEauthorblockA{\\ \IEEEauthorrefmark{2}Airbus Defence and Space, Stevenage, United Kingdom
		\\\{sevket.cetinsel, robert.ro.hughes\}@airbus.com }
	\IEEEauthorblockA{\\ \IEEEauthorrefmark{3}Applied DSP and VLSI Research Group, University of Westminster, London, United Kingdom,
		\\kalei@westminster.ac.uk}
}
\date{}

\maketitle

\begin{abstract}
This paper presents a two stage approach for the processing of frequency-stacked mobile sub-bands. The frequency stacking is performed in the analogue domain to enable the use of a wideband ADC, instead of employing multiple narrowband ADCs, to support multiple antenna elements for digital satellite beamforming. This analogue front end provides a common broadband digital interface to the on-board processor and can be configured to support multiple satellite missions, reducing the cost of commissioning a digital processor for individual satellite missions. This paper proposes a framework on the specification of digital prototype filter for the Analysis of frequency-stacked mobile sub-bands. The computational complexity of the Analysis operation, with two digital filter alternatives, are evaluated. A series of results, taken from our European Space Agency sponsored project, are presented here to demonstrate the applicability of the proposed two stage approach, reporting on the savings in power consumption when an Nth-band all-pass based recursive filter having an Infinite Impulse Response is used as the digital prototype filter.    
\end{abstract}


\IEEEpeerreviewmaketitle

\section{Introduction}

\begin{figure}[!t]
\centering
        \includegraphics[scale=0.75, trim=60.0mm 30.0mm 90.0mm 64mm, clip]
 {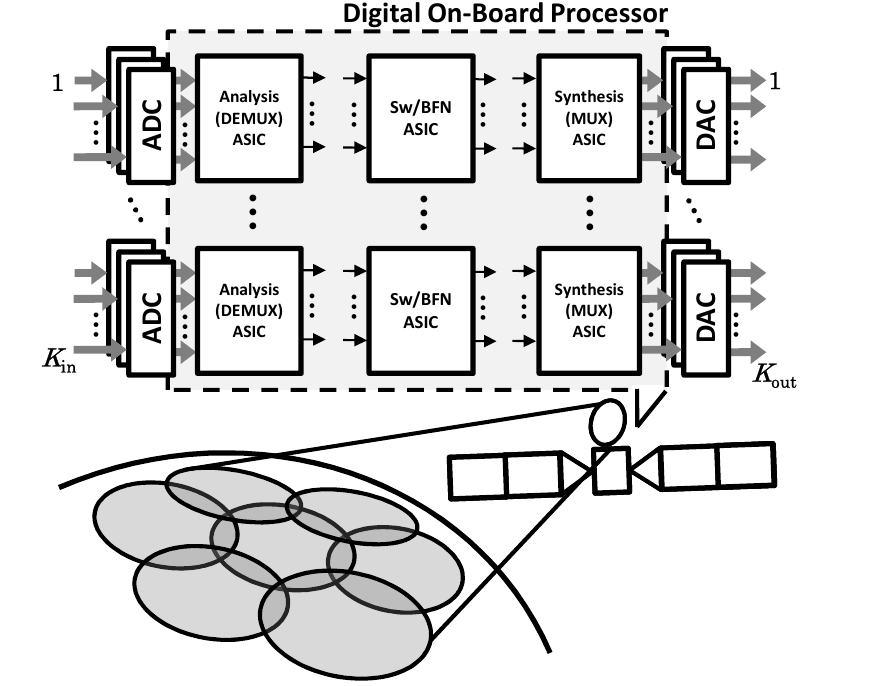}
	\caption{A representation of a beamforming satellite generating multiple spot beams and equipped with a digital OBP. The digital OBP contains three sets of ASICs for the Analysis, routing and the Synthesis of the communication spectrum as well as delivering the switching and the beamforming operations. A total of $K_{in}$ and $K_{out}$ ports are used at the input and the output of the OBP respectively, each served by several data conversion units.}
	\label{fig1}
\end{figure}

\IEEEPARstart{B}{eamforming} satellites make it possible to easily realise frequency reuse in the regions the satellite is operational. They provide flexibility and improved efficiency in the use of the available frequency spectrum. The on-board beamforming approach enhances satellite based broadband and high-speed communications, which is at the same time a potential key technology for the delivery of 5G and future wireless communication standards \cite{Piero2020}. Very High Throughput Satellites (VHTS), as well as satellites providing Mobile Satellite Services (MSS), are equipped with multiple antenna elements to facilitate beamforming both for the transmit and receive modes of operations. The signals collected from multi-beam systems are treated by a number of signal processing units and finally converted to the digital domain using multiple Analogue-to-Digital Converters (ADCs) to enable the use of a digital On-Board Processor (OBP) for the channelisation and baseband processing of the communication spectrum. An illustrative example is shown in Fig.1 for a transparent processing satellite, where the primary operations of the digital OBP are assigned to three sets of Application-Specific Integrated Circuits (ASICs) that are; (1) the Analysis (sometimes referred to as the DEMUltipleXer (DEMUX)),  (2) Switch/BeamForming Network (Sw/BFN) and the (3) Synthesis (sometimes referred to as the MUltipleXer (MUX)) \cite{AirbusProcessor2016} \cite{AirbusProcessor2007}. Here, the Analysis ASICs are responsible for the division of the frequency modulated spectrum into multiple bands, while the reconstruction of the spectrum is delivered by the Synthesis ASICs. The cross-connections deployed across the Sw/BFN ASICs simultaneously give full routing flexibility as well as delivering the digital beamforming operations. Finally the Digital-to-Analogue Converters (DACs) take the processed signals back to the analogue domain.

\begin{figure*}[!t]
	\includegraphics[scale=0.7, trim=30.0mm 55.0mm 30.0mm 50mm, clip]{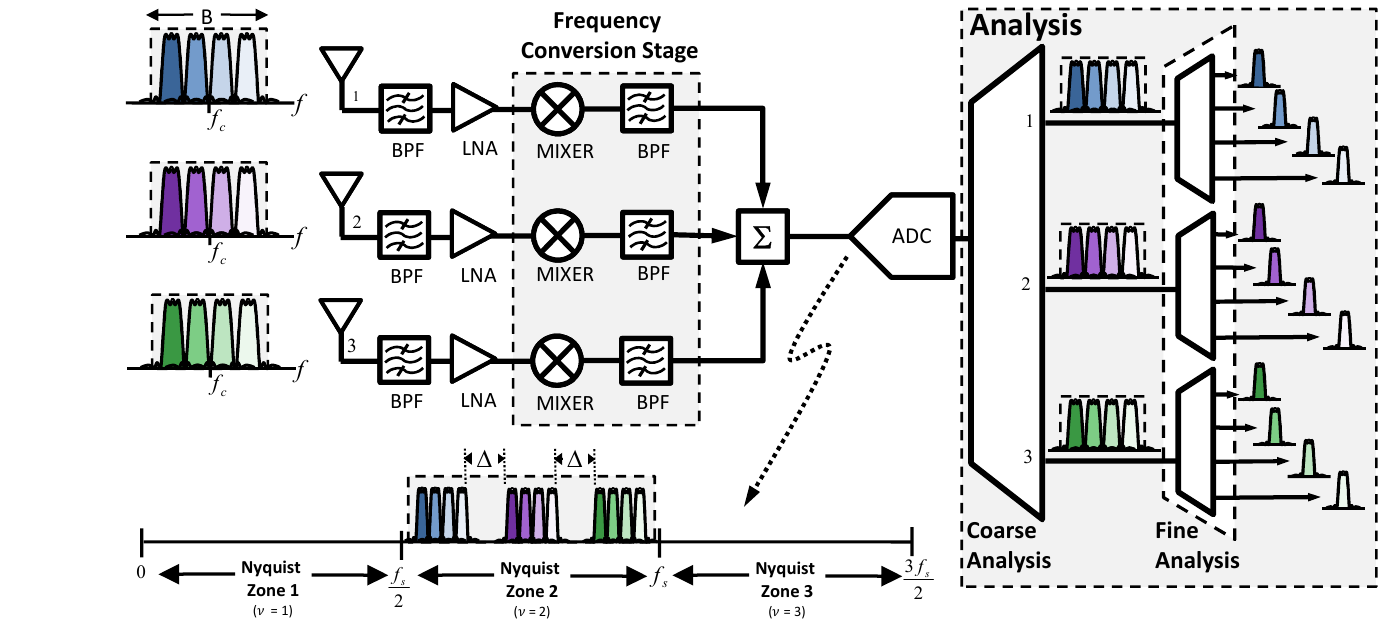}
	\caption{Frequency stacking scheme and the signal processing operations associated with this scheme both in analogue and digital domains. In the analogue domain, the signals from multiple antenna elements are combined and then processed by a single ADC unit. In the digital domain the two stage Analysis operation first demultiplexes the frequency-stacked sub-bands and then further decomposes each sub-band into multiple user channels for beamforming and sub-band processing operations.}
	\label{fig2}
\end{figure*}

\begin{figure}[t]
	\centering
	\includegraphics[scale=0.53, trim=65.0mm 45.0mm 65.0mm 48mm, clip]{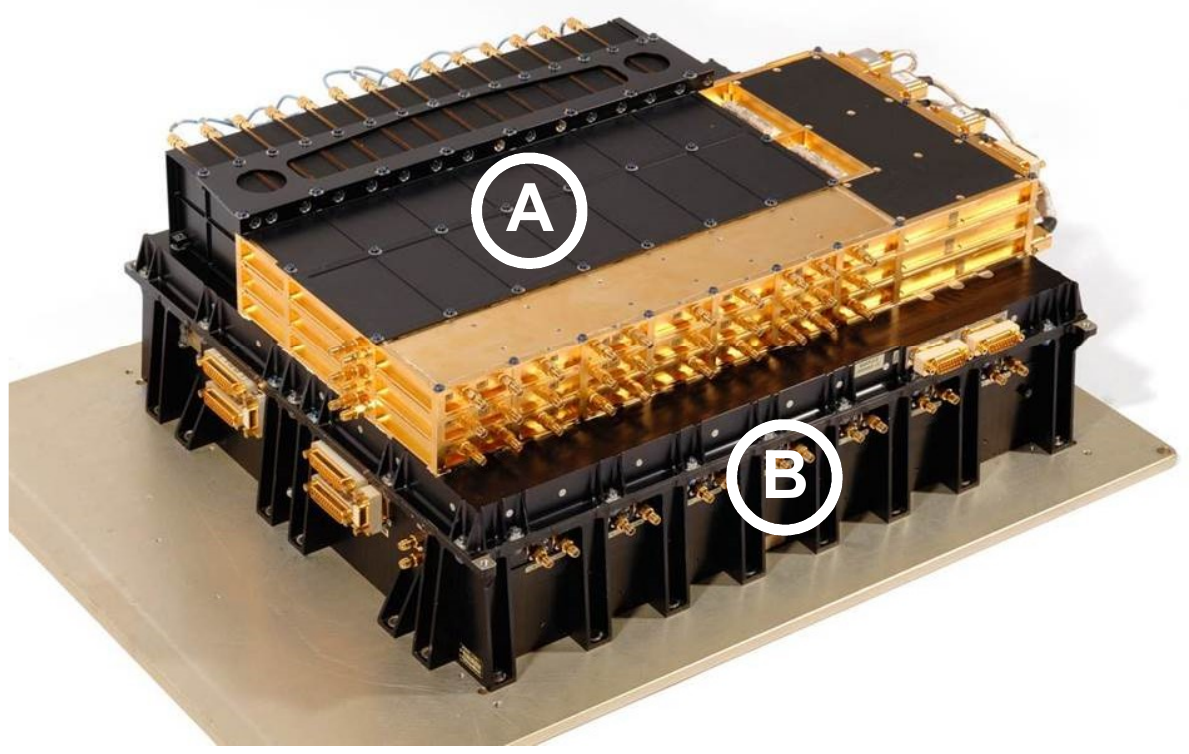}
	\caption{The photo of the processing hardware (presented here with permission of Airbus), with the (A) RF and (B) digital processors integrated in a single unit . The frequency stacking and the digital OBP operations are delivered by the integrated processor developed under the Alphasat programme.}
	\label{fig1a}
\end{figure}

As the processing bandwidths of the ADCs are increasing with the continuing technological advances, it is possible to combine the signals from multiple antenna elements and reduce the number of ADCs used on-board \cite{AirbusPatent01} \cite{AirbusPaper}. A representative scenario is shown in Fig.2, where a single ADC is shared by multiple antenna elements. Here the aim is to reduce the cost of the on-board processing by reducing the number of processing chains for data conversion and to provide a common broadband digital interface to the digital OBP. The model shown in Fig.2 is simplified for a better visualisation of the concept presented in this paper, where the Frequency Division Multiplexed (FDM) signals, received by the three antenna elements, are stacked in the frequency domain using a combination of analogue pre-processing units, composed of a mixer and a Band Pass Filter (BPF), operating at different mixing frequencies for each branch. This operation will be referred to as ``Frequency Stacking'' throughout the rest of our paper. Another important aspect is which Nyquist zone ($\nu$) the stacked sub-bands are to be located and this plays an important role in the choice of the mixing frequencies.

\add{The frequency stacking solution may be advantageous in comparison to configurations where an ADC per antenna element with direct digitisation is used. This is particularly because}
\begin{itemize}
\item \add{The improvements in ADC technologies over the last two decades have delivered broadband ADCs with similar performance and power as previous generations of narrowband ADCs, while the development of moderate bandwidth ADCs has not followed the same path. }
\item \add{In principle, developing such narrowband ADCs, customised for our specific needs (e.g. on a mixed signal ASIC), would increase the number of RF interfaces on each chip by a factor in proportional to the number of antenna feeds, which may not be possible to accommodate.}
\item \add{Furthermore, it is worth noting that the cost of hi-reliability mixed-signal components is a main driver in the overall cost of the onboard processor, and the reduction of the number of converters results advantageous both in terms of costs, integration, and overall flexibility to accommodate different missions.}
\end{itemize}

\add{On the other hand, comparing the complexity of frequency stacking solution with alternative configurations using an ADC per antenna element for direct digitisation;}
\begin{itemize}
\item \add{Direct digitisation is often unfeasible due to the high gain required between the antenna port and the ADC. In this respect frequency conversion is beneficial in reducing amplifiers oscillations by splitting the gains between different frequencies.}
\item \add{The ADC design is largely driven by signal RF frequency both in terms of ADC sampling clock jitter and analogue bandwidth (while ADC sampling rate may be limited to signal bandwidth).}
\item \add{Band-limiting is required to limit out-of-band interference and aliasing effects.}
\end{itemize}

\add{Due to the these aspects, it is often preferable to consider configurations where before digitisation the antenna element signal is first filtered, amplified (LNA) and converted (with the downconverter including further amplification and filtering). The net result is that the complexity of the analogue stage of the frequency stacking configuration is comparable to that of per-element down-converter+ADC configurations.}

The integrated processing equipment, developed under ESA's Alphasat programme \cite{Alphasat}, is an example to the hardware implementation of the concept \add{described so far} \remove{presented}, which is shown in Fig.3. The pre-processing to deliver the frequency stacking operations takes place within the RF processor (which is the top section labelled as (A)), while the digital OBP  (the bottom section labelled as (B)) is used for the baseband operations \cite{AirbusNGP} shown in Fig.1.

Once the resulting signal from the frequency stacking is converted to the digital domain, using a single wide-band ADC, the three operations listed above (Analysis, Sw/BFN and Synthesis) can now be performed by the digital OBP.
For the \change{delivery}{implementation} of the Analysis and Synthesis operations
\add{several options exist. Per-channel methods (i.e. the realization of a single digital filter per channel) becomes rapidly inefficient when the number of sub-bands increases. To reduce complexity and power consumption, multirate and fast-transform based methods have been developed, including:}
\begin{itemize}
\item \add{Hierarchical Multi-Stage (HMM) {\protect{\cite{Gockler_1988}}}. The demultiplexer features a modular design, with a binary tree structure, which allows a bandwidth hierarchy in power of 2.}
\item \add{Fast-transform modulated filter banks {\protect{\cite{Bellanger_Daguet_1974}}}-{\protect{\cite{Heller_et_al_1999}}}, which allow very efficient demultiplexing of fixed-width filtering channels (larger channels can be obtained via contiguous multiplexing of individual sub-channels).}
\end{itemize}
A real or a complex modulated filter bank is a cost efficient choice, which requires a digital prototype filter and a discrete transform such as the Discrete Fourier Transform (DFT) or the Discrete Cosine Transform (DCT) \cite{Aero01}, \cite{Aero02}, \cite{Mishra}. The computational complexity of such filter banks depend on the number of sub-bands to be supported by the digital OBP and the design requirements associated with the digital prototype filter to obtain the lowest power/order for a given set of satellite communication requirements and power budget. To achieve a low complexity alternative, in this paper an all-pass based recursive prototype filter with its use for the frequency-stacked mobile sub-bands have been investigated and will be reported.

\add{The main unique contributions of this article can be summarised as follows;}
\begin{itemize}
\item \add{Providing a comprehensive discussion on the frequency-stacking scheme for on-board processors, for the first time in the literature to the best of our knowledge, a use case for the Nth-band all-pass based recursive filters to support the Beamforming operation for communication satellites is disclosed.}
\item \add{in Section II.B, a novel way to evaluate the computational complexity of the Analysis operation to demultiplex the frequency-stacked sub-bands using an IIR filter candidate is presented. To the best of the authors$'$ knowledge, there is no other mathematical equation that can be used for the estimation of the number of filter coefficients to implement a Nth-band all-pass based recursive filter for an Almost Linear Phase (ALP) response.}
\item \add{in Section III.A, a two-step framework is proposed to specify prototype filters for the Analysis of frequency-stacked mobile sub-bands applicable to wideband ADCs.}
\end{itemize}

During our ``REconfigurableFiLter-banks of Efficient Channelisation for Satellites'' (REFLECS) project \cite{ReflecsPaper}, sponsored under the European Space Agency’s (ESA’s) ARTES Advance Technology Programme, we accommodated a digital prototype filter with an Infinite Impulse Response (IIR) and ALP, in the form of an Nth-Band all-pass based recursive filter, for the Analysis and the Synthesis of frequency-stacked sub-bands. This resulted in noteworthy savings, in particular a reduction in the region of 50\% in power dissipation for the Analysis of the frequency-stacked sub-bands, at the same time meeting the requirements on the allowable phase distortion.

In the next section we shall provide more details on the frequency stacking scheme and the system model of the proposed Analysis with two prototype filter alternatives. Section III discusses and formulates a framework on how to establish digital prototype filter specifications for the Analysis of the frequency-stacked mobile sub-bands, presented along with the results and observations on prototype filter realisations from the REFLECS project. The final section presents the concluding remarks from our findings.

\section{Frequency Stacking \remove{of Mobile Sub-Bands}}
\add{The main purpose of onboard processors is to enable flexible management of the available payload resources according to the traffic distribution. In conventional frequency-reused multibeam systems, the onboard processor provides beam-to-beam interconnectivity, where each uplink beam channel is divided in several sub-channels that can be individually routed to downlink beams. Considering that the beam to beam interconnection is typically realised in the frequency domain, the architecture is also generically known as Sub-channel-Switched or Satellite-Switched FDMA (SS-FDMA).
More recently payload systems have been proposed to exploit advanced Massive Multi-Input Multi-Output (M-MIMO) techniques to push to the limit frequency reuse {\protect{\cite{Marzetta_2015}}}-{\protect{\cite{Angeletti_De_Gaudenzi_2021}}}}.

\add{A reference payload model assumes separated forward (FWD) and return (RTN) processors (Fig.~1). Forward repeater creates the path from the gateway beams to the user beams and viceversa for return repeater. No beamforming is foreseen on the feeder-link side, while several possible antenna configurations {\protect{\cite{Angeletti_et_al_2003}}} with different beamforming needs to be considered on the user link:}

\begin{itemize} 
\item\add{Single Feed per Beam (SFPB) - No beamforming.}
\item\add{Array Fed Reflector (AFR) - Beamforming with limited number of feeds-per-beam (typically 7 to 20) and partial feeds sharing among beams.}
\item\add{Multi-Matrix Array Fed Reflector - Beamforming similar to AFR with the addition of a stage of appropriately connected multiport amplifiers (i.e. cascade of Input Network - INET, Amplifiers, Output Network - ONET).}
\item\add{Imaging Phased Array (IPA) - Full sharing of feeds per beams.}
\item\add{Direct Radiating Array (DRA) - Full sharing of feeds per beams.}
\end{itemize}

\add{The main effects of the different antenna configurations will be on the form of the array response to users in different far-field locations. With respect to the described configuration order, locality and amplitude variability will be higher for the top configurations while it will become of almost constant amplitude with phase only variation for the lower configurations.}
\add{According to the localisation effect, the channel bandwidth and amplitude level may largely vary among different element channels, nevertheless, to preserve the capability of performing beamforming in the digital domain, all the converters must work in constant gain mode and the frequency staking must account for the constant user bandwidth per each element channel.}

Because the development times and costs for radiation hardened ASICs are very high, in practice it is not viable to design a new ASIC specific for each satellite mission. A single solution that is applicable to all of the communication bands is highly desirable. Such a solution increases the applicability of a single development to a wider range of missions, and thereby reduces the overall cost of the OBP design and implementation for satellite telecommunications. A commonly proposed solution is to use different analogue pre-processing units to chop up wide signal bands or stack narrow ones, which is the main motivation behind the use of the frequency stacking scheme. 

\add{Frequency staking exploits the now common paradigm in software radio of employing wideband A/D and D/A converters, as they can accommodate a larger variability of the channel configurations. This entails that the converter intrinsically work in multicarrier, a condition that is already unavoidable for two main reasons:}
\begin{itemize}
\item \add{in the user uplink, the satellite antenna sees a multitude of active users with several terminals contemporarily active in the same bandwidth;}
\item \add{in the user downlink, the signal for each radiating element is a composite mix of the signal to be transmitted to different users in visibility of the antenna.}
\end{itemize}
\add{In consequence, the further multicarrier condition introduced by the frequency stacking is not particularly detrimental on the level of the quantization noise on the individual channels. Nevertheless it requires that the payload signal level budgets and the dimensioning of the pre-processor and post-processor gains are adequately taken into account in the overall design (including processing gain due to the small improvement due to the fact that a small portion of the quantization noise falls within the useful channel bands, and the beamforming processing gain).}

For the processing of the frequency-stacked sub-bands in the digital domain, in this paper we present a two-stage approach. The first stage is the \textit{Coarse Analysis}, which divides the frequency-stacked input spectrum into equal coarse channels to retrieve the individual sub-bands received by each antenna element. In other words, the first stage is to revert back the frequency stacking operation digitally. The second stage is the \textit{Fine Analysis}, where the purpose is to extract individual user channels.
The number of user channels \add{in the user bandwidth and their modulation format} depends on the mobile communication standard \add{(e.g. GMR-1 {\protect{\cite{ETSI_GMR_1}}}, GMR-2 {\protect{\cite{ETSI_GMR_2}}}, DVB-SH {\protect{\cite{DVB-SH}}} or the most recent 5G standards for Non-terrestrial-Networks {\protect{\cite{NTN_17}}}) and can include FDM, TDM, CDM and OFDM as well as their multiple access variants.} \remove{or alternatively could be customised to meet customer needs.}
\add{In this respect while the Fine Analysis and Synthesis stages are more strictly interlinked to the air interface standard, the Coarse Analysis and Synthesis stages are largely independent on it and are mainly dependant on the frequency stacking scheme.}
\add{Additionally, for all those architectures, where the user channels occupy all the user-bandwidth (e.g. M-MIMO and beam-hopping systems {\protect{\cite{Angeletti_et_al_2006}}}), the Fine Analysis-Synthesis stage becomes completely transparent.}
Therefore, authoring the design requirements for the Fine Analysis is relatively more straightforward. This makes the formulation of the Coarse Analysis a more interesting research problem, which we shall address in Section III. The Synthesis of the signals back for re-transmission is also performed in two stages which is presented within the same section too. All of these operations need for a Digital Signal Processing (DSP) architecture designed following the \textit{uniform modulated filter bank theory}, which will be looked at in the next sub-section.

\subsection{System Model for the Analysis}
The uniform modulated filter bank architecture is composed of; an Analysis Filter Bank (AFB), to separate the communication spectrum, $X(z)$, into multiple sub-bands, and a Synthesis Filter Bank (SFB), that multiplexes the same number of frequency bands back together forming a single output stream, $Y(z)$, as shown in Fig.4(a). Note that the AFB and the SFB are the underlying DSP operations for the Analysis and the Synthesis ASICs respectively, introduced in Fig.1.

The ordering of the SFB and the AFB may also be interchanged to implement a transmultiplexer \cite{Johan01} to convey a frequency multiplexed signal over a wireless communication channel as shown in Fig.4(b). The top figure in Fig 3.(a) outlines the digital processing units associated with the AFB (which is known as the \textit{DFT modulated filter bank} due to the DFT modulating the filters), where $N$ is the number of sub-bands that the spectrum of $X(z)$ is to be decomposed into. The AFB involves downsamplers and unit delayors (each delayor is shown with a triagle ($\triangledown$) in Fig.4(a)), realised as a \textit{commutator} when combined together. Additionally, the polyphase sub-filters,  $H_n\left(z\right)$, and the Inverse Discrete Fourier Transform (IDFT) unit are the two other DSP blocks associated with the DFT modulated filter bank. The internal structure of the SFB is very similar with that of the AFB, the difference being a simple reversal of the order of operations as well as the IDFT being replaced with DFT. 

\begin{figure}[t!]
	\centering
	\begin{subfigure}{}
  \includegraphics[scale=0.52, trim=60.0mm 30.0mm 10.0mm 24mm, clip]
  {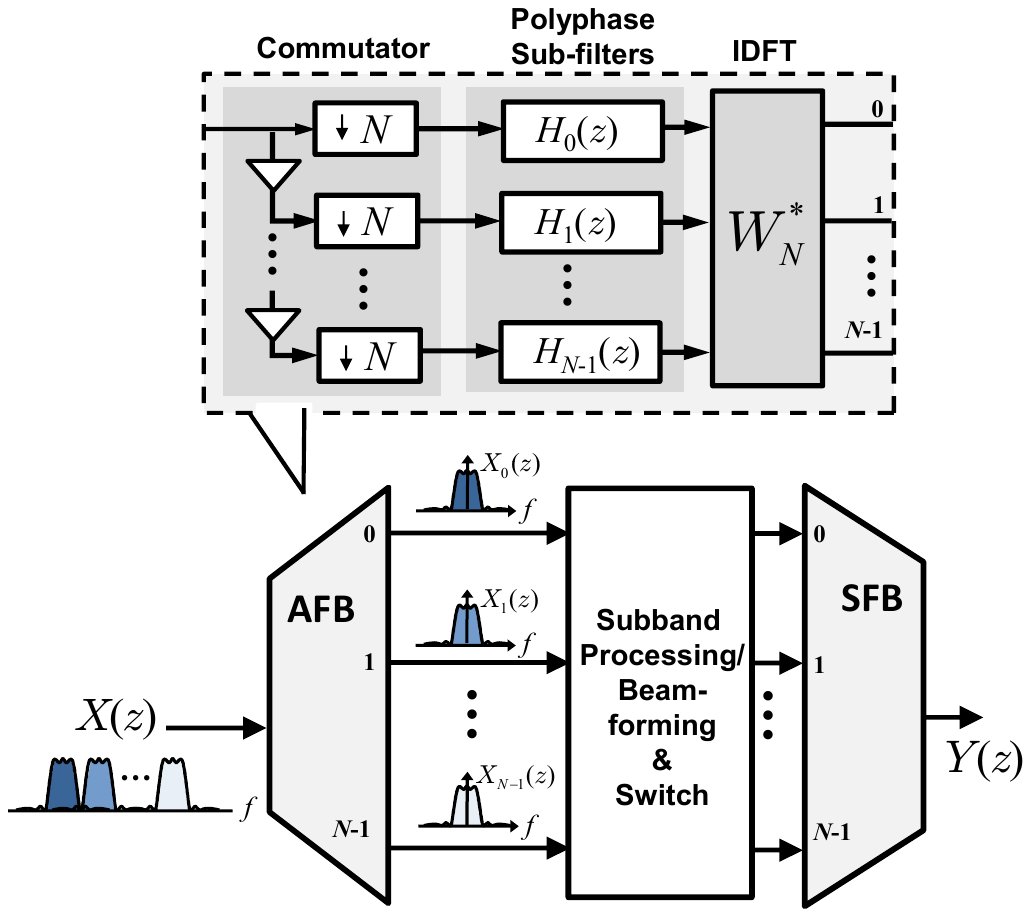}
		\\(a)
	\end{subfigure}
	
	\begin{subfigure}{}
  \includegraphics[scale=0.55, trim=70.0mm 60.0mm 20.0mm 64mm, clip]
  {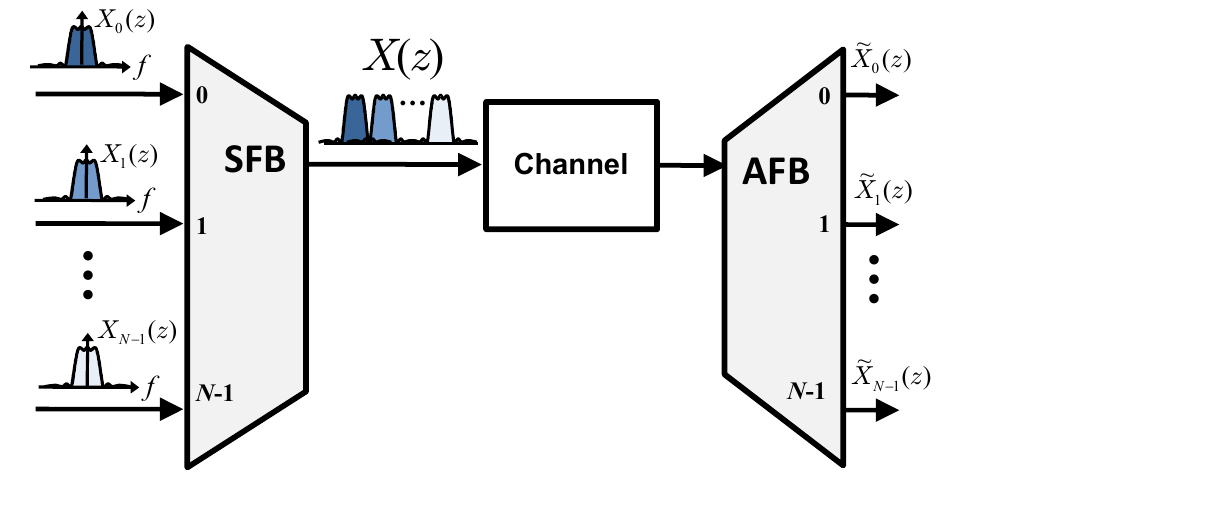}
		\\(b)
	\end{subfigure}
	\caption{(a) The Analysis Filter Bank (AFB) and the Synthesis Filter Bank (SFB) to analyse and synthesise a communication spectrum (b) The same can be achieved in the form of a Transmultiplexer, where the aim is to convey the message through a wireless communication channel with minimal distortion and therefore to obtain an estimate of each sub-band at the output of the AFB, i.e. ${\tilde{X}}_i(z)\approx X_i(z)$.}
	\label{fig3}
\end{figure}

Using the DFT modulated filter bank, shown in Fig.4(a), the output of the AFB corresponding to the $l$-th sub-band can be formulated as follows

\begin{equation} \label{GrindEQ__1_} 
\begin{split}
X_l\left(z\right)
&=\frac{1}{N}\sum^{N-1}_{n=0}{H_n\left(z\right)}\left\{\sum^{N-1}_{m=0}{W^{-n\left(m+l\right)}_Nz^{-\frac{n}{N}}X\left(z^{\frac{1}{N}}{\ W}^m_N\right)}\right\} \\ 
&= \frac{1}{N}\sum^{N-1}_{n=0}{H_n\left(z\right)G_{n,l}(z),} 
\end{split}
\end{equation} 

\noindent where ${\ W}_N=\ e^{-j\frac{2\pi }{N}}$  is the complex exponential kernel associated with the IDFT. 

Knowing that $G_{n,l}\left(z\right)$ is a function of the input signal and parameters that are fixed for a given number of sub-bands; the performance, accuracy and the complexity of the Analysis is controlled primarily by the polyphase sub-filters $H_n(z)$ for $n$=1,\dots ,\textit{N}-1, which are the polyphase components of the lowpass prototype filter, $H(z)$, with transfer function

\begin{equation} \label{GrindEQ__2_} 
H\left(z\right)=\ \sum^{N-1}_{n=0}{H_n\left(z^N\right)\ z^{-n}}.\  
\end{equation} 

\begin{figure*}[b]
	\begin{equation} \label{GrindEQ__3_} 
	\begin{split}	 \hline
	H_n\left(z\right)=\ \frac{1}{N}\prod^{n_{FoS}-1}_{m=0}{\overbrace{\frac{{\alpha }_{n,m}+z^{-1}}{1+{\alpha }_{n,m}z^{-1}}}^{ \begin{array}{c}
			First\ order\ \\ 
			Section \end{array}
		}\ } =\frac{1}{N}\times \frac{b_{n,0}+b_{n,1}z^{-1}+\dots +b_{n,\ n_{FoS}-1}z^{-{(n}_{FoS}-1)}+z^{-n_{FoS}}\ }{1+b_{n,\ n_{FoS}-1}z^{-1}+\dots +b_{n,1}z^{-(n_{FoS}-1)}+b_{n,0}z^{-n_{FoS}}\ }.
	\end{split}
	\end{equation} 
\end{figure*}

It is possible to realise $H(z)$ as a recursive filter with an Infinite Impulse Response (IIR) or alternatively a non-recursive filter with a Finite Impulse Response (FIR). The two will be discussed in the following two sub-sections as the lowpass prototype filter alternatives of the DFT modulated filter bank.  \\

\begin{figure}[t]
	\centering
	\includegraphics[scale=0.85,trim=0.0mm 5.0mm 0.0mm 10mm, clip]{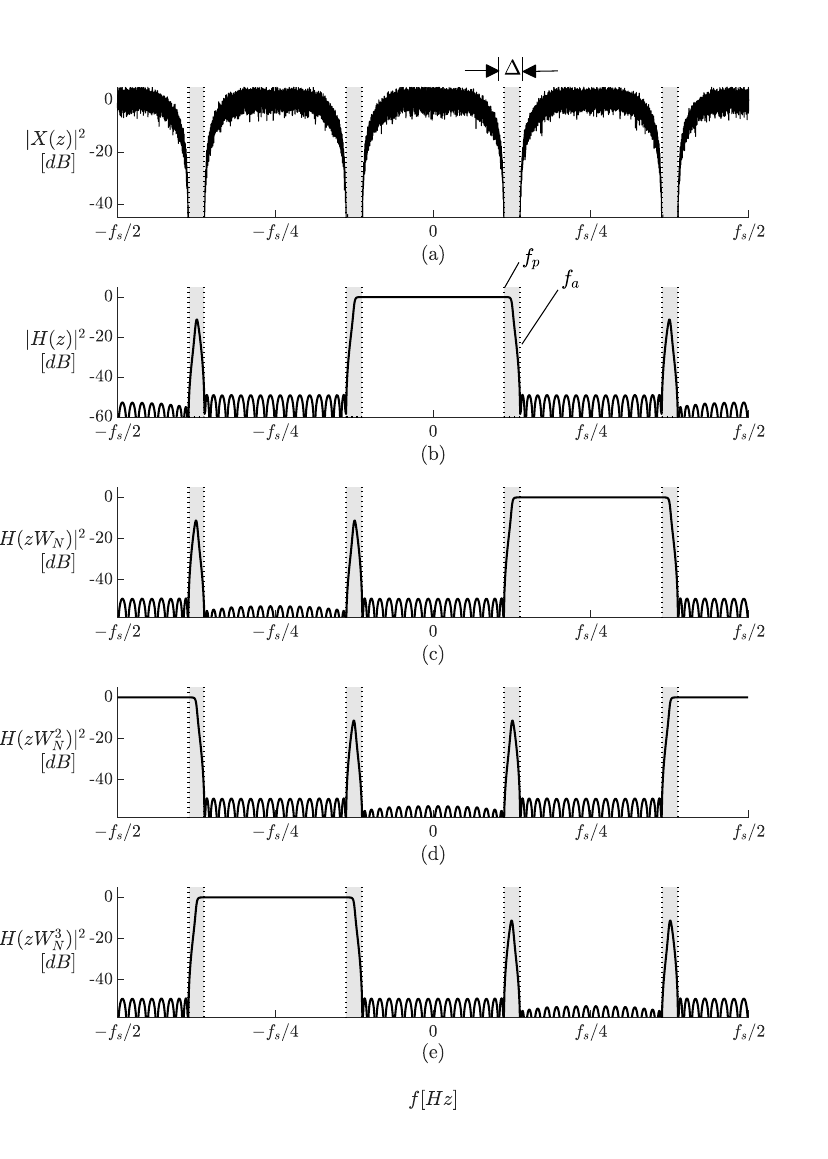}
	\caption{(a) Spectrum of $N=4$ sub-bands separated by guardbands. (b) The magnitude response of the prototype lowpass filter (it’s three DFT modulated copies are shown in (c), (d) and (e)).}
	\label{fig4}
\end{figure}

\noindent \textbf{Recursive Prototype Filter with an Infinite Impulse Response: } The first alternative to realising $H(z)$ is based on an IIR response, for which the transfer functions of the polyphase sub-filters are obtained as the product of all-pass First Order Sections (FoS) as in \eqref{GrindEQ__3_}.

\add{For the low-level realisation of these filters, special attention has to be paid considering all possible low-level alternatives}. It is possible to combine the first order all-pass sections and implement $H_n\left(z\right)$ as a cascade of higher order sections \add{(second order being the most popular)} or alternatively \add{forming a} \remove{as} ${n_{FoS}}^{th}$ order IIR all-pass \add{filter} \remove{section} using \add{the} \remove{simple} first order section constructs. It should be noted that the latter case is known to adversely impact the coefficient quantisation and necessitates for more complex arithmetic operators and higher datapath wordlength. \add{On the other hand, if second order sections are used, critical nodes have to be identified within all-pass sections and further decision have to be made {\protect{\cite{AllPass01}}} {\protect{\cite{AllPass02}}} to avoid unnecessary wordlength growths that would limit the performance due to feedback.} In \eqref{GrindEQ__3_},  $n_{FoS}$  is the number of first order sections per polyphase sub-filter, and ${\alpha}_{n,m}$ is the $m$th all-pass filter coefficient of the $n$th polyhase sub-filter, where at the same time ${-\alpha}_{n,m}$ is a pole of the IIR transfer function. On the other hand, $b_{n,m}$ represents the transfer function coefficients of  $H_n\left(z\right)$.

The total number of all-pass filter coefficients affects the complexity of the DFT modulated filter bank that will be looked at Section III. Considering \eqref{GrindEQ__2_} and \eqref{GrindEQ__3_}, the total number of filter coefficients to implement $H\left(z\right)$, using the recursive prototype filter with the IIR response, is  $L_{IIR}=N\times \ n_{FoS}$. Note that the scalar multiplication $1/N$ in \eqref{GrindEQ__3_} is not taken into consideration as a coefficient, which is simply for gain scaling and can be compensated if needed at a later stage of the processing chain within the OBP.

The lowpass prototype filter, with its polyphase sub-filters implemented as in \eqref{GrindEQ__3_}, is also known as the Nth-Band all-pass based recursive filter with IIR, which is known to provide a low complexity alternative to non-recursive digital filters. An almost linear phase with a practically perfectly flat magnitude response (microdB ripple) is achieavable, setting the first branch to a pure delay line, i.e. $H_0\left(z\right)=z^{-n_{FoS}}$, with ${\alpha }_{0,m}=0$ for $\forall $$\ m$. 

The most critical problem employing an Nth-Band all-pass based recursive filter for communication systems appears to be the undesirable spectral spikes present in the stopband region resulting from the imperfect cancellations in the transition regions. To avoid these spectral spikes that cause aliasing, the frequency-stacked sub-bands must be separated with guardbands \cite{Mathias}. A simple filtering scenario of four users is shown in Fig.5. In Fig.5(a), $X(z)=Z\{x(k)\}$ is the frequency-stacked signal at the input to the Nth-band all-pass based recursive filter (\textit{k} being the discrete time index and $Z\{.\}$ is the z-transform) and a certain frequency spacing is required between the stacked sub-bands in the form of a guardband represented with $\Delta$. The magnitude response of the prototype lowpass filter (shown in Fig.5(b)) and it's three DFT modulated copies (in Fig.5(c), (d) and (e)) are deployed in a way that, their spectral spikes and the transition bands overlap with the spectral portions of $X(z)$, where no useful signal is present (shown with grey shaded regions in Fig.5). It is possible to accommodate up to $N$ sub-bands (for a maximally decimated filter bank) where the $n$-th user should be located in the frequency region $\left\{\frac{f_s}{N}n-f_p,\frac{f_s}{N}n+f_p\right\}$, as shown in Fig.5 (a)-to-(e), where $f_p$  is the passband edge of the prototype filter and $f_s$  is the sampling rate.  The frequency stacking operation should be delivered in a way such that no signal of interest is present between neighbouring communication bands, which is ensured by the analog BPFs placed after the mixing stage in Fig.2. At the minimum, $\Delta$ can be set equal to the transition bandwidth, $f_t$, of the prototype filter so that $\Delta=f_t=f_a-f_p$, where $f_a$ is the stopband edge of the prototype filter. Based on this, one can calculate the percentage of the available ADC bandwidth occupied by the guardbands in terms of  the width of the $f_t$ with respect to the amount of the spectrum occupied by $N$ number of frequency-stacked sub-bands, as follows,  

\begin{equation} \label{GrindEQ__4_} 
\Delta_{\%}={\Delta}f\times N\times 100\ [\%], 
\end{equation} 

\noindent as a percentage, where ${\Delta}f=f_t/f_s$ is the normalised transition bandwidth. \\

\noindent \textbf{Non-Recursive prototype filter with a Finite Impulse Response:} A non-recursive prototype filter with an FIR response has traditionally been the most popular choice for satellite OBPs designed for telecommunication applications \cite{Sulli}. It is also possible to formulate the transfer function of the polyphase sub-filters for an FIR response using first order sections, similar to \eqref{GrindEQ__3_}, as follows

\begin{equation} \label{GrindEQ__5_} 
\begin{split}
H_n\left(z\right) 
&=\ b_{n,0}\prod^{n_{FoS}-1}_{m=0}{1+{\alpha }_{n,m}z^{-1}\ } \\
&=b_{n,0}+b_{n,1}z^{-1}+\dots +b_{n,\ n_{FoS}}z^{-n_{FoS}} 
\end{split}
\end{equation} 

\noindent where ${-\alpha }_{n,m}$  is the $m$th zero of the nth sub-filter. For the FIR, it is not necessary to divide a long transfer function into first or second order sections unlike the IIR. However, to write both \eqref{GrindEQ__3_} and \eqref{GrindEQ__5_} in a similar manner as well as to use  $n_{FoS}$  in the transfer function of the FIR too, the transfer function of the FIR is written as a product of first order sections too.  Based on \eqref{GrindEQ__2_} and \eqref{GrindEQ__5_}, the total number of filter coefficients that implements $H\left(z\right)$ is $L_{FIR}=N\times (n_{FoS}+1)$ for the non-recursive prototype filter alternative with FIR response.

In this paper the transfer functions, and therefore the filter coefficients, of the lowpass prototype filters are obtained through the well-known Remez-type algorithms using Matlab. For the IIR case, the algorithmic steps from \cite[Chapter 14]{SaramakiBook}, \cite[Chapter 1]{LinPhase01} and \cite[Chapter 11]{Harris02} can be followed when obtaining filter coefficients to implement $H_n(z)$ to achieve an almost-linear phase response, while for the FIR we have chosen minimum order equiripple implementation to be able to achieve the lowest filter order as a low complexity competitor to the IIR prototype filter alternative.  In the next section the computational complexity of these two prototype filter alternatives are evaluated and compared. 

\subsection{Computational Complexity of the recursive and non-recursive prototype filter alternatives}

In this section, we evaluate the computational complexity of the Analysis to demultiplex the frequency-stacked sub-bands, considering the two prototype filter alternatives formulated in the previous section. \add{The intention here is to disclose a novel and at the same time a practical way to the complexity assessment, keeping contributions from low-level hardware realisations out of the scope. For this reason, t}he evaluation is performed in terms of the number of real additions and real multiplications, which is a widely used metric to assess the complexity of DFT modulated filter banks \cite[pg.51]{Haanthesis}. Additionally, in the complexity estimation below, we have selected to replace a complex multiplication by the 3 real multiplications and 5 real additions \cite{3mul}. \\

\noindent \textbf{Candidate 1: Finite Impulse Response (FIR) filter candidate:} Our first candidate is based on the non-recursive prototype filter alternative and named the ``FIR filter candidate'' due to its FIR response. The complexity of an FIR filter is well studied and several alternative equations were proposed to quantify the number of filter coefficients as a function of the passband ripple, ${\delta }_p$, stoppband ripple, ${\delta }_s$, and the transition bandwidth, $f_t$. Although more accurate equations are available in the literature, estimating the total number of coefficients for an equiripple FIR filter, we will prefer to use a more practical equation \cite[pg 106]{Taylor}

\begin{equation} \label{GrindEQ__6_} 
L_{FIR}\cong \frac{0.0714\ \times (-10{log}_{10}\left({\delta }_p\times {\delta }_s\right)-15)}{\Delta {\rm f}}.       
\end{equation} 

Assuming the signal at the input of the filter, $x(k)$, to be complex valued (which leads to a complex by real multiplication at the digital filtering stage, as the coefficients of the FIR filter are real valued), the number of real additions to be performed to realise the FIR prototype filter is,

\begin{equation} \label{GrindEQ__7_} 
a_{FIR}=2\times (L_{FIR}-N),      
\end{equation} 

\noindent while the number of real multiplications is 

\begin{equation} \label{GrindEQ__8_} 
p_{FIR}=2\times L_{FIR}.          
\end{equation} 

To deliver the IDFT, Inverse Fast Fourier Transform (IFFT) is used. The computational complexity of the IFFT operation is 

\begin{equation} \label{GrindEQ__9_} 
a_{IFFT}=9\times \frac{N}{2}{log}_2(N)\textrm{, and }p_{IFFT}=3\times \frac{N}{2}{log}_2(N),
\end{equation}

\noindent where $a_{IFFT}$  is the number of real additions and $p_{IFFT}$  is the number of real multiplications associated with the IFFT operation. In \eqref{GrindEQ__10_}, the $\frac{N}{2}{log}_2(N)$ term is the well-known complexity parameter used to estimate the order of the computational load of the FFT/IFFT. Because an FFT/IFFT butterfly circuit is made up of 2 complex-by-complex additions (due to the use of 1 subtractor and 1 adder, which is equal to 4 real additions) and 1 complex-by-complex multiplier, which is equivalent to 3 real multipliers and 5 real adders, the total number of real additions for a single butterfly circuit is 9 while the number of real multiplications is 3. This is reflected in the formulation of \eqref{GrindEQ__9_}. 

The total number of additions and the multiplications needed to realise Candidate 1, based on \eqref{GrindEQ__7_}, \eqref{GrindEQ__8_} and \eqref{GrindEQ__9_} is

\begin{equation} \label{GrindEQ__10_} 
{a_1=\ a}_{FIR}+\ a_{IFFT}=2L_{FIR}+\frac{9N}{2}{log}_2(N)-2N 
\end{equation}
 
\noindent and

\begin{equation} \label{GrindEQ__11_} 
{p_1=\ p}_{FIR}+\ p_{IFFT}=2L_{FIR}+\frac{3N}{2}{log}_2(N).
\end{equation}

\begin{figure}[t]
	\centering
        \includegraphics[scale=1,trim=5.0mm 0.0mm 0.0mm 0mm, clip]{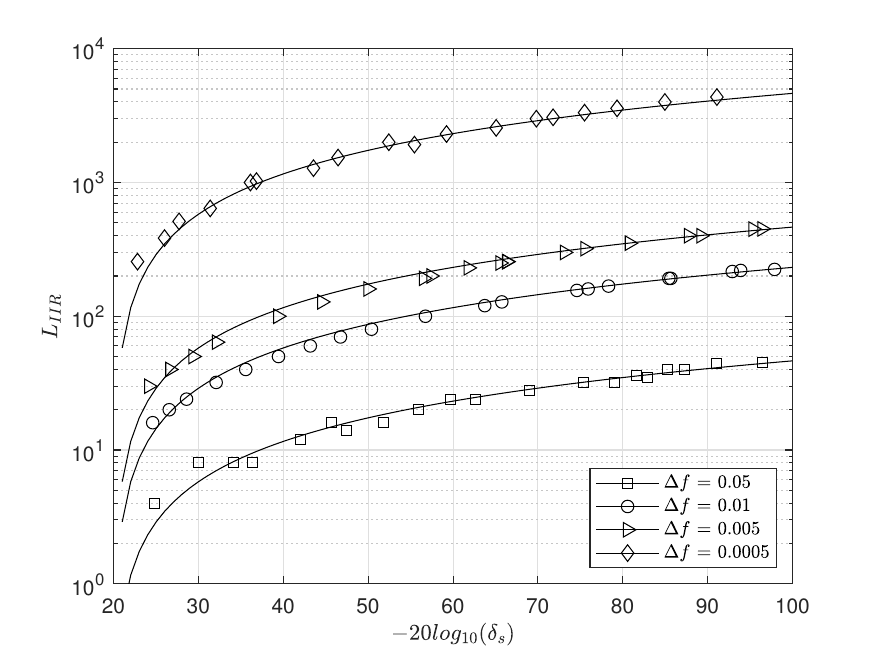}
	\caption{The change in the number of all-pass filter coefficients with respect to the change in the stopband attenuation and the normalised transition bandwidth for Nth-band All-pass based recursive filter with ALP response.}
	\label{fig5}
\end{figure}

\noindent \textbf{Candidate 2: Infinite Impulse Response (IIR) filter candidate:} Our second candidate is based on the recursive prototype filter alternative and named the ``IIR filter candidate'' due to its IIR response. \remove{To the best of the authors knowledge, there is no mathematical equation that can be used for a quick estimation of the number of filter coefficients needed to implement a Nth-band all-pass based recursive filter for an ALP response. Therefore} \add{For this purpose}, we conducted a set of simulations to come up with an empirical equation, similar to \eqref{GrindEQ__6_}, to compare the complexity of implementing the IIR filter candidate against that of FIR. We designed several recursive filters by varying the stopband ripple  ${\delta }_s$ and $f_t$, and recorded the number of first order all-pass sections, which is equivalent to $L_{IIR}$ delivering an ALP response. The result of these simulations can be seen in Fig.6. The simulation results have shown that $L_{IIR}$ can be approximated as a function of the stopband ripple and the transition bandwidth in our formulation as follows

\begin{figure}[t]
	\centering
        \includegraphics[scale=0.8,trim=0.0mm 5.0mm 0.0mm 10mm, clip]{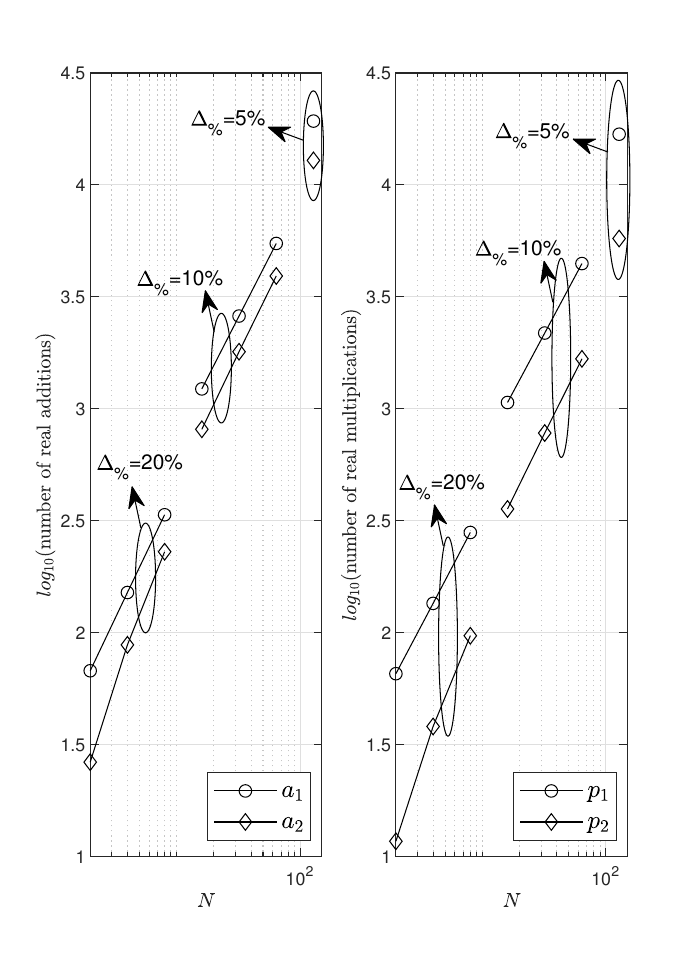}
	\caption{The computational complexity of the Analysis evaluated for both FIR and IIR filter candidates. The evaluation is performed in terms of the number of real additions and real multiplications.}
	\label{fig6}
\end{figure}

\begin{equation} \label{GrindEQ__12_} 
L_{IIR}\cong \frac{0.058\times (-10{log}_{10}\left({\delta }_s\right)-10)}{\Delta {\rm f}}.      
\end{equation} 

In Fig.6, the markers indicate the result from the simulations, while the lines with no markers are plotted using \eqref{GrindEQ__12_}. As can be seen, the lines closely follow the markers, which justifies the usability of \eqref{GrindEQ__12_} for our complexity estimation. To avoid any additional complexities in the design of the filters, we limited the simulations for the number of channels to 2$\leq$\textit{N}$\leq$128, and the guardbands to occupy $5\%\le \Delta_{{\rm \%}}\le 40\%$. Because the passband ripples are so small (in the region of $\mu$dBs) and it does not make a noteworthy contribution to the filter complexity, ${\delta }_p$ is not used in \eqref{GrindEQ__12_}.  

Bearing in mind, a first order all-pass section contains a multiplier and two adders (so that $a_{IIR}$  is double the $p_{IIR}$) and excluding the first branch of the Almost Linear Phase Nth-band recursive filter, which is a pure delay line and does not contribute to the number of arithmetic operations, the total number of real additions and the multiplications of Candidate 2 can be written as follow

\begin{equation} \label{GrindEQ__15_} 
{a_2=\ a}_{IIR}+\ a_{IFFT}=4\left(1-\frac{1}{N}\right)L_{IIR}+\ \frac{9N}{2}{log}_2(N) 
\end{equation} 

\noindent and

\begin{equation} \label{GrindEQ__16_} 
{p_2=\ p}_{IIR}+\ p_{IFFT}=2\left(1-\frac{1}{N}\right)L_{IIR}+\frac{3N}{2}{log}_2(N).
\end{equation} 

To compare the computational complexity of the two candidates, Fig.7 is generated using \eqref{GrindEQ__10_} and \eqref{GrindEQ__15_} for calculating the number of real additions, \eqref{GrindEQ__11_} and \eqref{GrindEQ__16_} for real multiplications. Fig.7 is generated for $N$ being a power-of-two to be able to use the derived equations and it shows that, as expected, the IIR filter candidate outperforms the FIR filter candidate providing a lower complexity alternative with a lower number of real multiplications and additions in all cases realised in Fig.7. Generating Fig.7, the $\Delta_{\%}$  is reduced as $N$ is increased for a more realistic utilisation of guardbands.

\section{Implementation of Frequency-Stacking Scheme using Nth-Band recursive filter}

\begin{figure}[t!]
	\centering
	\begin{subfigure}{}
		\includegraphics[scale=0.5, trim=30.0mm 60.0mm 15.0mm 44mm, clip]{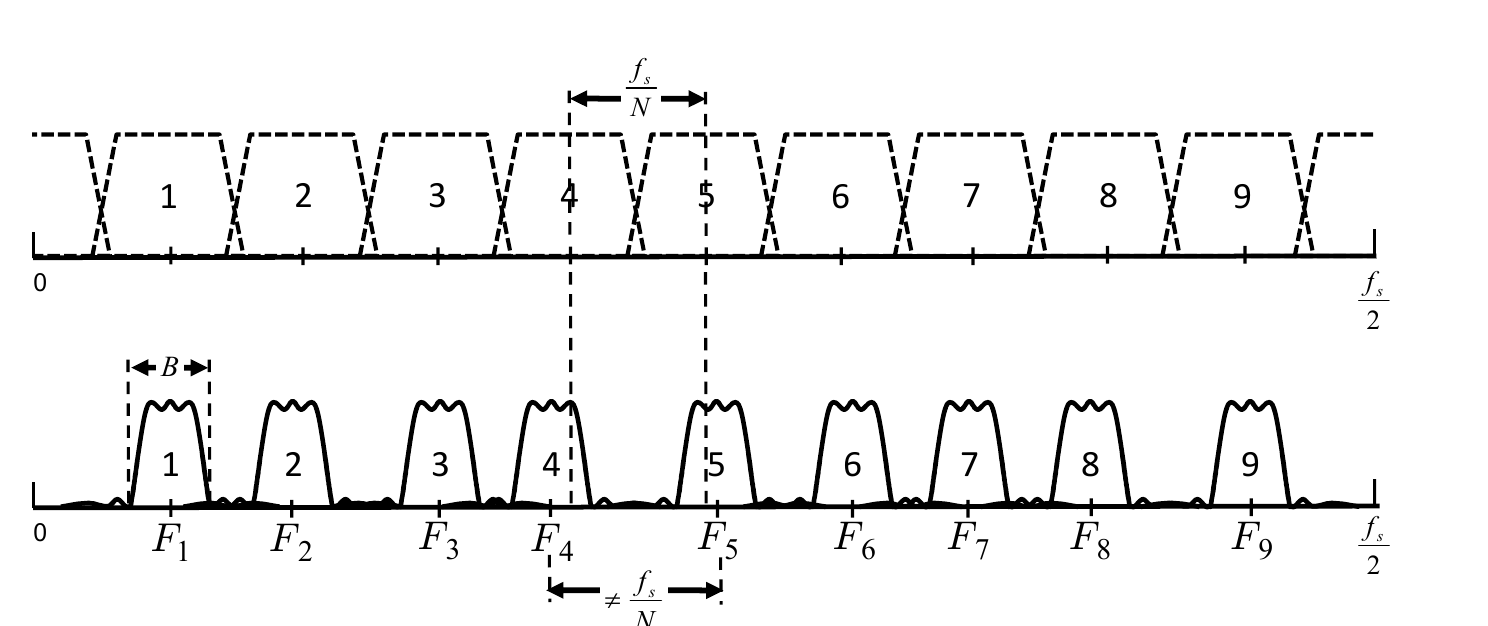}
		\\(a)
	\end{subfigure}
	
	\begin{subfigure}{}
		\includegraphics[scale=0.5, trim=15.0mm 60.0mm 20.0mm 54mm, clip]{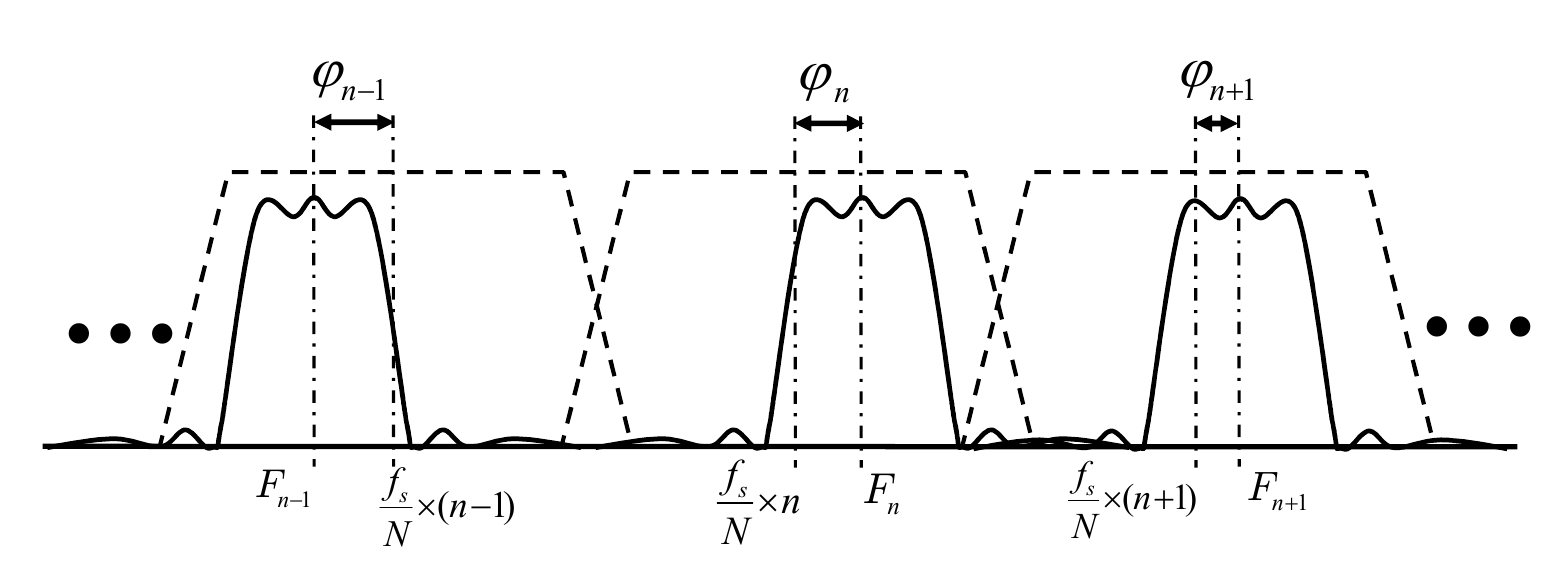}
		\\(b)
	\end{subfigure}
	\caption{(a) The frequency range of \{0, $f_s$/2\} is divided into 10 equal bands. The top figure shows the DFT modulated filters, while the bottom figure is the frequency-stacked mobile sub-band signals. Pay attention to the non-uniform spacing between individual sub-bands. Here, sub-bands corresponding the DC and the Nyquist frequencies intentionally left unused. (b) The value of ${\varphi }_n$ is the deviation of the centre frequencies of the DFT modulated filters from their corresponding mobile sub-band signals.}
	\label{fig7}
\end{figure}

The structure illustrated in Fig.2 has two important implications. Firstly, it must be recognised that the elemental signals will be combined to form steerable beams and must accurately preserve their relative amplitude and phase information in order to support the beamforming process. The local oscillator signals driving the mixers in the frequency conversion stage are all required to maintain a constant phase relationship and must in practice be generated from a common master reference oscillator signal. In this implementation, the master oscillator frequency is $f_o$ =$10$ MHz from which all other local oscillators are derived. Therefore, all of the mixing frequencies used for frequency stacking must be an integer multiple of $f_o$, i.e.  ${\beta }_n\times f_o$, where ${\beta }_n\in {{\mathbb{Z}}}^+$.

Secondly, it must be possible to precisely line up channel boundaries with the edges of the frequency-stacked mobile sub-bands. The digital prototype filter must be sized to accommodate the frequency allocations for the services which are to be processed. These differ for a number of parameters including the region on the Earth to be served or the mobile communication standard to be used. In this paper we consider Europe (Region 1) L-Band MSS uplink spectrum 1626.5 to 1675 MHz (including the extension band), which is the widest frequency allocation of interest for MSS. This means, each mobile sub-band has the same bandwidth of $B$=48.5 MHz and the centre frequency of each mobile sub-band is $f_c$ = 1650.75 MHz. 

A further major technology constraint comes from the choice of the space-qualified ADCs and DACs. The ADC in particular often represents the single most important limitation on achievable signal performance. For this study, as a broadband ADC, we considered EV12AS200 device from e2v \cite{e2v}. This is a 12-bit component, sampling at rates of up to 1.5 GHz and dissipating 3.2 W. In line with these specifications, a sampling clock frequency of $f_s$=1280 MHz has been assumed for our frequency stacking process.  

\begin{table}[t]
	\caption{A Framework to Specify the Lowpass Prototype Filter to the Analysis of Frequency-Stacked Sub-bands}
	\begin{center}
	\begin{tabular}{cl} 
		\hline \vspace{0.1cm} \\ 
		\ & \textbf{\textit{Input}} $f_s$, $f_o$,$\ f_c$, $\nu$, B \textit{and} 	$N=2{\times N}_c$ \vspace{0.1cm}\\ 
		1: & $\rho$=$\left\lfloor \frac{\nu }{2}\right\rfloor $$;$ s=sgn(2$\rho$-$\nu$+1/2)  \vspace{0.2cm} \\
		2: &\textbf{\textit{for}} $n$ = 1,\dots ,$\ N/2$-1 \textbf{\textit{do}}\\ 
		3: & \hspace{0.5cm} ${\varphi }_n={\mathop{argmin}_{{\beta }_n\in \mathbb{Z}^+} \left(\left|F_n-n\frac{f_s}{N}\right|\right)\ }$ \\                
		\ & \hspace{1cm} \textbf{\textit{where} }   $F_n=s\times ({\beta }_nf_o-\ \left(f_c-{\rho \ f}_s\right))$ \\ 
		4: &  \textbf{\textit{end}} \vspace{0.1cm}\\ 
		5: & $f_p={max \left\{\left[{\varphi}_1,{\varphi}_2,\ \dots ,\ {\varphi}_{N/2-1}\right]\right\}+\frac{B}{2}\ }$ \vspace{0.1cm}\\ 
		6: & $f_a=\frac{f_s}{N}-f_p$\\  \\ \hline 
	\end{tabular}
	\label{tab1}
	\end{center}
\end{table}

\subsection{Prototype Filter Specification for Frequency-Stacking Scheme}

For our implementation, we consider a total of 9 mobile sub-bands, each received by a different antenna element, and then all of them are combined in the frequency domain as a result of the frequency stacking operation. The frequency-stacked signal, digitised at the output of the selected wideband ADC, is real-valued and therefore the frequency range of \{0, $f_s$/2\} is divided into $N_c$=10 equal bands as shown in the top plot shown in Fig.8(a). Each sub-band is served by a DFT modulated copy of the prototype filter represented by the dashed lines in Fig.8. Here we intentionally avoid the utilisation of the DC and the Nyquist frequencies, keeping the remaining 9 bands occupied. To enable the use of a DFT modulated filter bank (similar to Fig.4(a)) the spectrum between \{-$f_s$/2, $f_s$/2\} has to be considered and therefore $N= 2N_c$. The bottom plot in Fig.8(a) demonstrates the spectrum of nine frequency-stacked sub-bands, where $F_n$, for $n$=1,\dots,$N$/2-1, represents their centre frequency in the 1st Nyquist zone, and there is no signal at the DC and at the Nyquist frequency.

We would like to draw attention to the fact that the frequency-stacked mobile sub-bands are not spaced evenly across the available Nyquist zone unlike the DFT modulated filters. This can be observed from Fig.8(a), comparing the position of the DFT modulated filters and the mobile sub-bands. While the spacing between adjacent DFT modulated filters is the same and it is  $\frac{f_s}{N}$ (see the top plot in Fig.8(a)), the frequency stacked sub-band signals are not necessarily spaced equally (see the bottom plot in Fig.8(a)).

In Table I, a two-step framework is proposed to specify the passband edge, $f_p$, and stopband edge, $f_a$, of the prototype filter to analyse the frequency-stacked mobile sub-bands (line 5-6), produced as a result of the frequency stacking scheme. In this framework, the first step is to calculate the deviation of $F_n$ from the centre of the DFT modulated filters for each mobile sub-band (line 2-4), and this deviation is denoted by ${\varphi}_n$. The value of ${\varphi}_n$ differs for different sub-bands (see Fig.8(b)) and the aim with this framework is to minimise ${\varphi}_n$ for each sub-band selecting ${\beta }_n$. This minimisation is important for a wider transition bandwidth, which reduces the computational complexity of the prototype filter as formulated in the Section II. Note that the passband edge of the prototype filter has to be wide enough to accommodate the largest deviation as shown in Fig.8(b). Another important point is the selection of the Nyquist Zone, $\nu$, to be used for the frequency stacking process (See Fig.2 on the use of $\nu$). Line-1 in Table I is used to incorporate $\nu$ in the selection of mixing frequencies.       

The proposed framework is executed for the communication scenario explained in this sub-section (for $N_c$=10) and the resulting values are presented in Table II \add{for the two filter candidates}. The down-converted signals are sampled at the 2nd Nyquist zone ($\nu$ =2) so that lower mixer frequencies can be used. The most important observation is the confirmation of the use of the IIR filter candidate (Candidate 2) that it is possible to lower the number of filter coefficients by 70\% instead of using the FIR filter candidate (Candidate 1). In our ESA sponsored activity, the translation of this into hardware complexity and power dissipation was undertaken that we shall disclose next. \add{In addition to these two filter candidates, in Table II the prototype filter specifications from} \cite{Sulli} \add{and} \cite{Aero01} \add{are presented too for comparison purposes. It should be noted that several technology parameters can affect the design of a prototype filter. Therefore, a one-to-one comparison is usually not possible between prototype filters. However, Table II shows that the Nth-Band all-pass based recursive filter (Candidate 2) requires the lowest number of coefficients comparing to FIR filter reported from similar studies despite having more stringent specifications.}

\begin{table*}[t]
	\caption{Prototype Filter Specifications}
	\begin{center}
		\begin{tabular}{|c|c|c|c|c|} 
			\hline 
			\textbf{} & \textbf{Candidate 1} & \textbf{Candidate 2} & \add{{\textbf{\protect{\cite{Sulli}}}}} & \add{{\textbf{\protect{\cite{Aero01}}}}} \\ \hline 
			Operating Frequency & \multicolumn{2}{|c|}{64 MHz ($f_s$/20)} & \add{$f_s$/64} & \add{264 kHz} \textsuperscript{i} \\ \hline 
			Passband Edge & \multicolumn{2}{|c|}{29 MHz} & \multirow{2}{*}{\add{${\Delta}f=$0.0016}} \textsuperscript{ii} & \multirow{2}{*}{\add{${\Delta}f=$0.05}} \\ \cline{1-3} 
			Stopband Edge & \multicolumn{2}{|c|}{35 MHz} & & \\ \hline 
			Deviation of the Passband Ripple & 0.0492 dB & 140 $\mu$dB & \add{0.0275 dB} \textsuperscript{ii} & \add{0.1 dB} \\ \hline 
			Stopband Attenuation & 51.42 dB & 49.09 dB & \add{56 dB} & \add{70 dB} \\ \hline 
			Number of filter coefficients & \textbf{$L_{FIR}=$}600 & \textbf{$L_{IIR}=$}180 & \add{2048} & \add{241} \\ \hline 
			Phase deviation  & Linear Phase & $<1^o$ & \add{Linear Phase} & \add{Linear Phase}\\ \hline 
			Sampling Clock Frequency ($f_s$) & \multicolumn{2}{|c|}{1280 MHz} & \add{1500 MHz} & \add{1.25 MHz} \\ \hline 
			{Spectrum of Interest} & \multicolumn{2}{|p{3.5cm}|}{1626.5-1675 MHz \newline Europe (Region 1) L-Band}
			& \multicolumn{1}{|p{3cm}|}{\add{Uplink: 28.5 GHz, Downlink: 19.95 GHz} \newline } & \multicolumn{1}{|p{3cm}|}{\add{960 $-$ 1164 MHz} \newline \add{L$-$band Digital Aeronautical Communication System (LDACS)}}\\ \hline 
		\end{tabular}
		\label{tab2}
\end{center}
\end{table*}

\subsection{Results from the REFLECS project}

The proposed two stage approach was modelled and tested as part of our REFLECS project, sponsored by the ESA \cite{ReflecsPaper}. During this project, the project team worked on a number of low-power DSP solutions applicable to narrowband (typically corresponding to MSS at L- or S-band), and broadband communication scenarios (which are typified by high throughput multi-beam services at Ku- and Ka-band). In this section only the findings from the investigated narrowband MSS scenario will be disclosed, which is relevant to the concepts presented in this paper. 

The  high-level model of the OBP developed during the REFLECS project, to test the end-to-end performance of a channeliser, is shown in Fig.9. The channeliser is based on our two stage approach, where the Coarse Analysis and Coarse Synthesis form the ``coarse channeliser'' and similarly the Fine Analysis and the Fine Synthesis form the ``fine channeliser''. Channel processing involves all of the baseband processing operations, digital beamforming and switching.

{\let\thefootnote\relax\footnotetext{\textsuperscript{i} A decimation factor of 2 was assumed. In \cite{Aero01}, the prototype filter was designed considering a passband narrower than 264 kHz.}}
{\let\thefootnote\relax\footnotetext{\textsuperscript{ii} Calculated based on the information that $\delta_p=\delta_s$ and Equation (5) given in \cite{Sulli}.}}

\begin{figure}[t]
	\centering
	\includegraphics[scale=0.75, trim=45.0mm 30.0mm 50.0mm 24mm, clip]
    {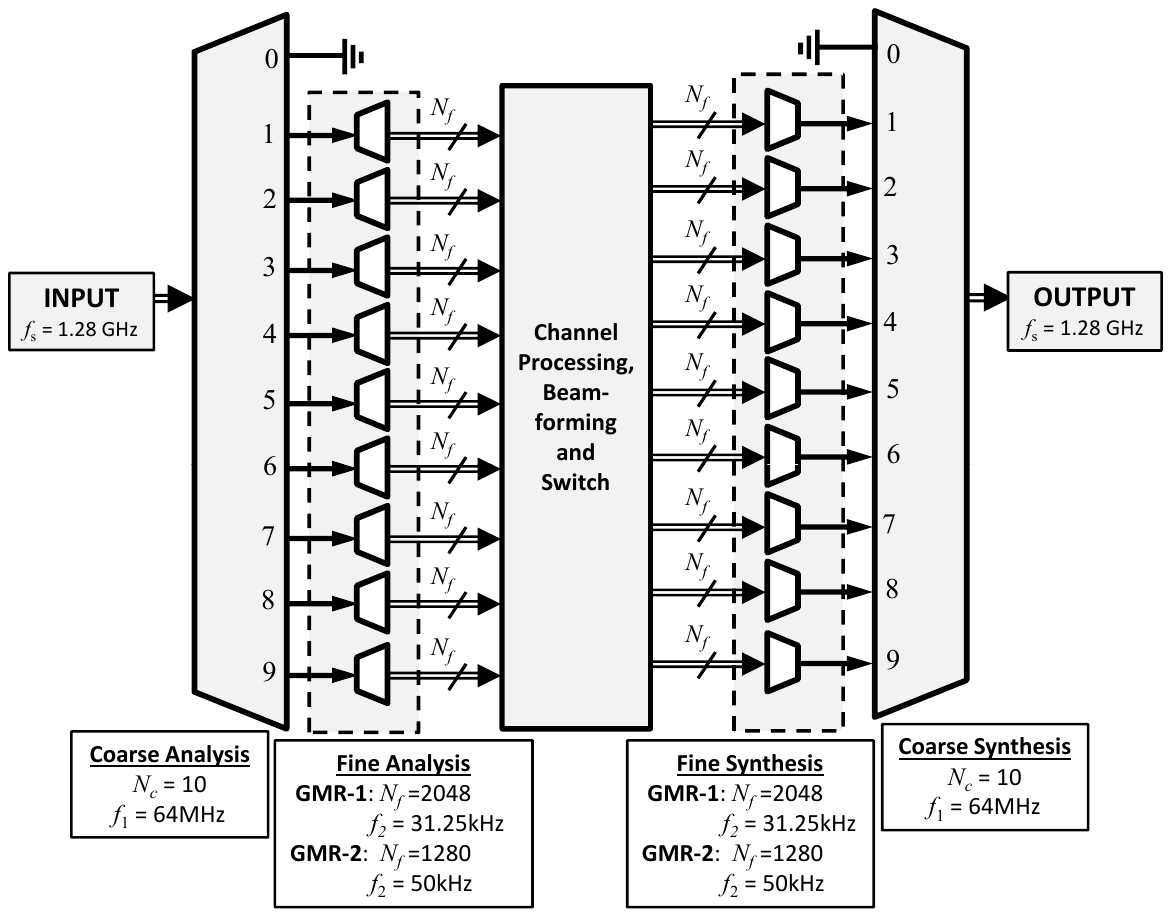}
	\caption{The high-level view of the digital OBP modelled to test the end-to-end performance of the channeliser based on the proposed two stage approach.}
	\label{fig8}
\end{figure}

\remove{The }MSS \add{satellites} typically follow frequency plans according to either the GMR-1 \add{\protect{\cite{ETSI_GMR_1}}} or GMR-2 \add{\protect{\cite{ETSI_GMR_2}}} standards. These are based on a grid of RF frequencies and differ in both the bandwidth (e.g. 31.25 kHz granularity for GMR1 or 50 kHz for GMR2) and edge frequencies. In the REFLECS project, the fine channeliser was developed to provide flexibility to support either scheme but not necessarily both at the same time. The fine channeliser permits precise tuning of grid edge frequencies to align with the GMR-1 or GMR-2 definitions in the L- and S- bands, with the guardband for each mobile user channel not exceeding 10\% of the channel bandwidth.

To comply with the specs of the wideband ADC, EV12AS200 \cite{e2v}, the input sampling rate was selected to be $f_s$=1280 MHz and the input signal in Fig.9 is real valued. The coarse channeliser was developed to process $N_c$=10 MSS sub-bands (similar to the model presented in Section III.A (see Fig.8)) stacked prior to analogue-to-digital conversion to increase the spectrum utilisation efficiency. Each output from the Coarse Analysis produces a signal at a rate of $f_1$=64 MHz. The frequency spectrum of the input to the Coarse Analysis for the GMR-2 test case is shown in Fig.10(a). Here, only nine sub-bands are occupied to intentionally avoid the use of DC and Nyquist frequencies. This is the reason why the $0^{th}$ output from the Coarse Analysis, shown in Fig 8, is terminated due to producing a `zero' output. In Fig.10(a) an inset plot that shows individual GMR-2 user channels from a portion of the spectrum is presented. 

\begin{figure}[t!]
	\centering
	\begin{subfigure}{}
		\includegraphics[scale=0.75, trim=5.0mm 0.0mm 0.0mm 0mm, clip]{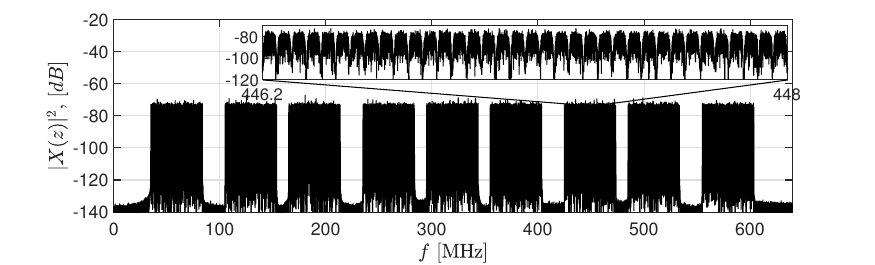}
		\\(a)
	\end{subfigure}
	
	\begin{subfigure}{}
		\includegraphics[scale=0.85, trim=5.0mm 0.0mm 0.0mm 5mm, clip]{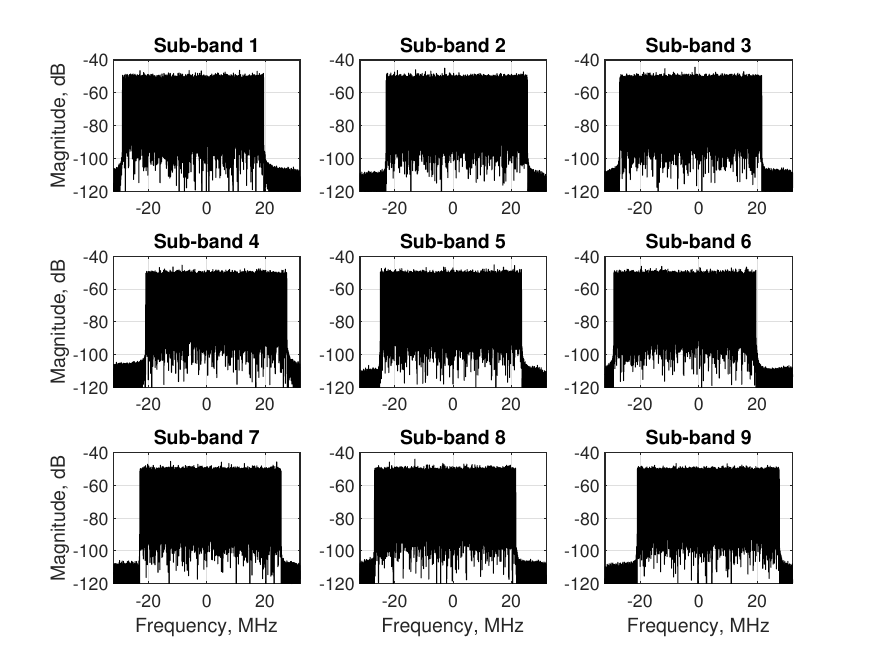}
		\\(b)
	\end{subfigure}
	\caption{(a) Nine frequency-stacked mobile sub-bands, 64 MHz each. The inset provides a close-up view to a portion of the spectrum to show how individual mobile user channels are frequency multiplexed for transmission based on the GMR-2 standard. (b) The output of the Corse Analysis, where all nine sub-bands are individually extracted.}
	\label{fig9}
\end{figure}

Using the proposed Coarse Analysis unit, it is possible to analyse and extract the nine frequency-stacked sub-bands as shown in Fig.10(b). Following this, the extracted sub-bands go into a secondary analysis operation to obtain much finer user channels, as part of the fine channeliser. For the fine channeliser, nine Fine Analysis units are needed, one per sub-band, in order to divide each subband into 31.25 kHz granularity for the GMR-1 or 50 kHz granularity for the GMR -2 standard. The number of user channels, $N_f$, at the output of each Fine Analysis unit is proportional to the user channel granularity, that is $N_f$ = 64MHz/31.25kHz = 2048 for GMR-1 and $N_f$ = 64MHz/50kHz = 1280 for GMR 2. An example to the 50 kHz user channels, successfully retrieved as a result of the Fine Analysis, is shown in Fig.11 for GMR-2. Similarly, the Coarse and Fine Synthesis units are to reverse the operations delivered by the Analysis units. Note that the DFT and IDFT within these channelisers are implemented using Fourier transform units that support sizes of non-power-of-two \cite{CoskunISCAS}, \cite{CoskunFlex}.

The power and hardware complexity of the proposed solution were evaluated using the activation rates and the number of primitive DSP components (i.e. adders, multipliers, register, RAM and I/O). Our evaluation has shown the OBP, based on the FIR filter candidate (Candidate 1), dissipates approximately 4.06 W power and uses 3.64 Mgates end-to-end including all stages of fine and coarse channelisers. Note that these figures were obtained when the design was evaluated  and optimised for power (with a trade-off in hardware complexity) and the overall structure includes the control logic and additional circuitry for reconfigurability if one or more sub-bands are not in use or switched-off. For the IIR filter candidate (Candidate 2), the power dissipation and the hardware usage, with the same assumptions, reduces to 3.13 W and 3.54 Mgates respectively providing 23\% saving in the power dissipation. The reduction in the power dissipation is more than 50\% (0.64 W versus 0.28 W for Candidate 1 and 2 respectively), if the Coarse Analysis is considered alone, which is the underlying DSP operation presented in this paper for the Analysis of the frequency-stacked mobile sub-bands. All our results reported here were obtained from bit-true hardware equivalent simulations giving a high degree of confidence in our findings. \add{Because the processor size and the number of the ASICs depend on the payload requirements (including the number of antenna elements in a beamforming satellite), the power figures and the gate count we provide in this study can be scaled to estimate the contribution of the Analysis for full OBP of another size.}

\begin{figure}[t]
	\centering
	\subfigure[][]{
		\label{gsm}
        \includegraphics[scale=0.5, trim=2.0mm 0.0mm 0.0mm 0mm, clip]{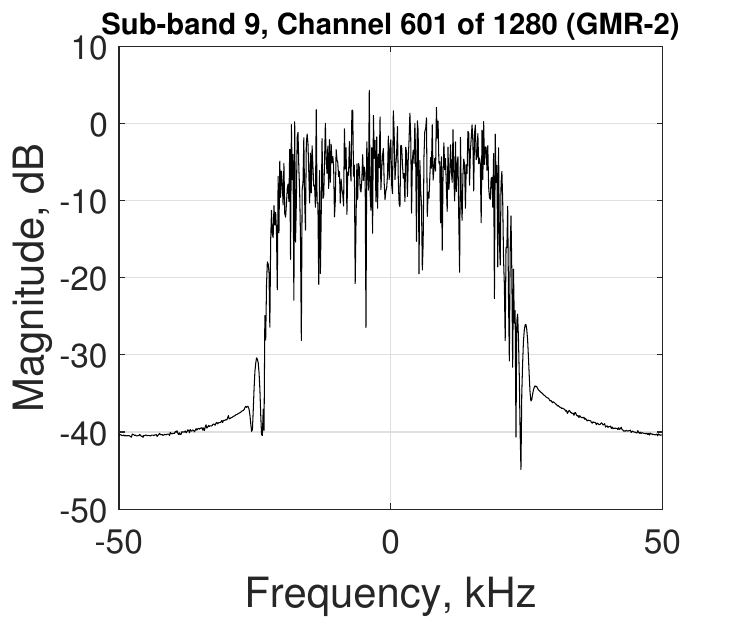} }
	\subfigure[][]{
		\label{gsm1}
        \includegraphics[scale=0.5, trim=2.0mm 0.0mm 0.0mm 0mm, clip]{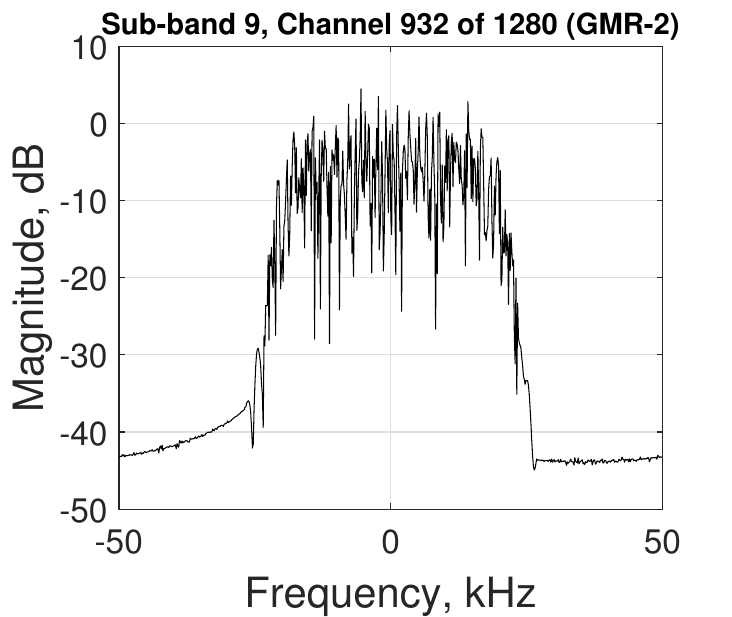} }
	\caption{It is possible to extract all 50 kHz mobile user channels ($N_c{\times}N_f$ = 12,800 GMR-2 channels in total) at the end of the Fine Analysis. Two of these channels are presented here which are randomly selected from the 9th sub-band.}
	\label{fig10}
\end{figure}

\begin{figure}[t]
	\centering
	\includegraphics[scale=0.75, trim=15.0mm 0.0mm 0.0mm 0mm, clip]{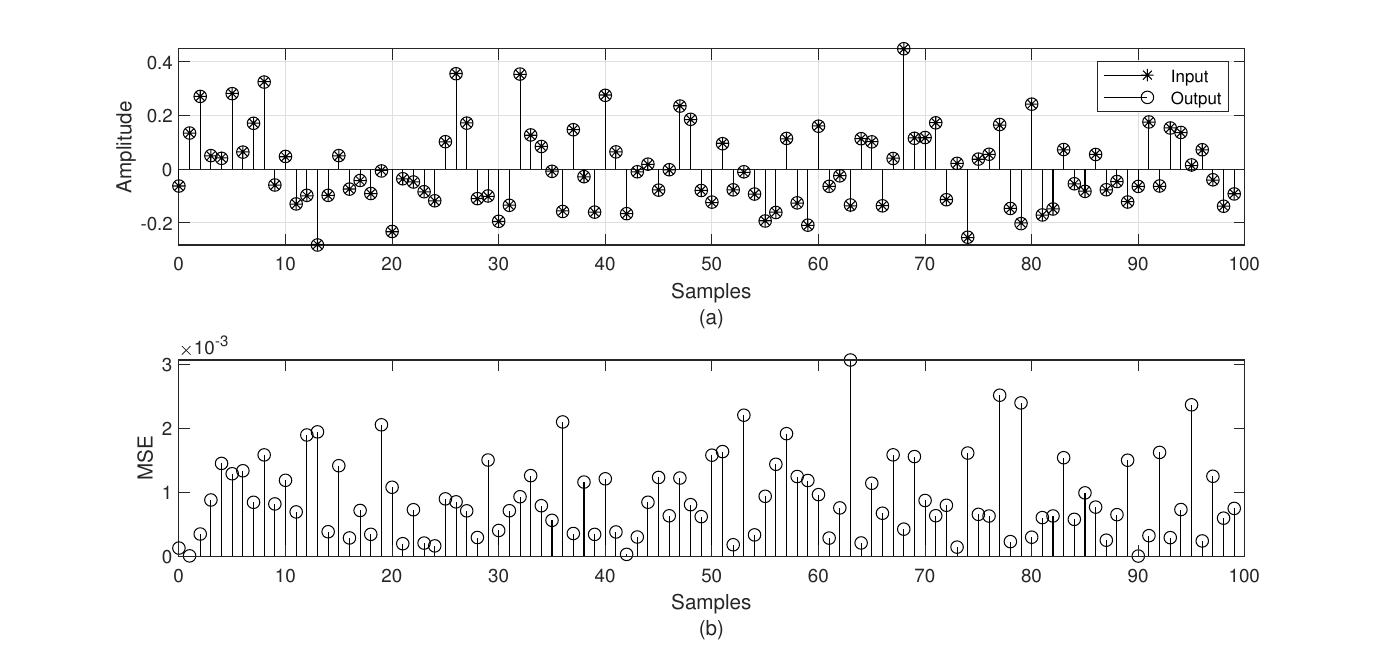}
	\caption{(a) ``Input'' and ``Output'' in the legend corresponds to the signal at the input and the output of the model presented in Fig.9. Only the first 100 samples are displayed. (b) End-to-End MSE performance of the OBP for GMR-2 data (when Coarse Channeliser is based on the IIR filter candidate).}
	\label{fig11}
\end{figure}

\begin{figure}[t]
	\centering
	\includegraphics[scale=1, trim=15.0mm 0.0mm 0.0mm 0mm, clip]{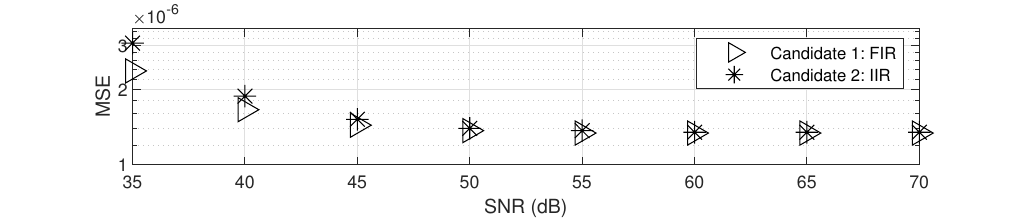}
	\caption{\add{End-to-End MSE performance of the OBP comparing two candidate filters under varying SNR conditions.}}
	\label{fig13}
\end{figure}

\add{In high throughput satellite applications, power efficiency trumps all other technical performance criteria. Therefore, for the REFLECS project, only ASIC realisations were considered as the power efficiency and radiation hardness of an ASIC implementation much outweigh that of an FPGA implementation in an equivalent technology node, even when taking into account the flexibility of the FPGA. The ASIC technology can offer higher power and area efficiency, due to their custom design, enabling a better utilisation of logic, which makes ASIC qualification a necessity for the qualification of such power hungry high throughput processors.}

%\add{For the REFLECS project only an ASIC realisations were investigated primarily due to the limited availability of radiation-hardened FPGAs to support high throughput tasks. Where massive amount of data is expected to be processed, commercially %available radiation-hardened FPGA technologies offer only a small proportion of what an ASIC can offer. In addition to this, ASIC technology can provide higher power and area efficiency, due to their custom design, enabling a better utilisation of %logic. There are recent developments from NanoXplore to support high capacity services using radiation-hardened FPGAs {\protect{\cite{NGUltra}}} and for a future study these devices may be considered to undertake certain high throughput tasks.}

To evaluate the error rate performance of the channeliser, the channel processing block in Fig.9 is disabled to be able to compare the input and the output to see how much distortion is introduced throughout the processor due to quantisation noise and aliasing. Fig.12(a) shows the input and output data sequences in and out of the  communication processor and Fig.12(b) is the difference between the input and the output in the Mean Squared Error (MSE) sense, when the IIR filter candidate is selected to be used at the coarse channeliser. The distortion is minimal, achieving an MSE of $1.35{\times}10^{-6}$, and that is not much different from the achievable MSE performance if the FIR filter candidate was used, which reaches an MSE rate of $1.29{\times}10^{-6}$. \add{The MSE performance was additionally investigated under varying SNR conditions changing the level of the Additive White Gaussian Noise (AWGN). The results are shown in Fig. 13. As can be seen here, in the MSE-sense the two candidate filters provide similar results too. In case of a higher noise level, observed between the stacked channels, the proposed IIR filter performs relatively worse than the FIR, which is due to the spectral spikes in the frequency response of the Nth-band all-pass based recursive filter. This is the primary reason why we considered the IIR candidate for the Coarse Analysis but not for the Fine Analysis. In Fine Analysis, we do not have full control on the signal that is received per mobile channel, while for Coarse Analysis, with the additional front end on-board, it is assured that there is no signal at these unwanted bands overlapping with the spectral spikes. The test in Fig.13 was conducted lowering the SNR down to 35 dB. It shall be noted that this is the SNR observed at the output of the analog filtering and if anything lower than 35 dB is expected then either the analog filters in the front end of the receiver shall be replaced to provide a better dynamic range or the IIR filter candidate shall be re-designed.} 
\section{Acknowledgement}
This work was funded by the European Space Agency under contract number 4000112150/15/NL/AD.

\section{Conclusion}
In this paper a two stage approach is presented to process frequency-stacked mobile sub-bands. We focused mainly on the first stage of this scheme where the Analysis of the frequency-stacked sub-bands is performed using a maximally decimated DFT modulated filter bank. For the lowpass prototype filter, the first candidate was to use a Linear-Phase FIR filter in order to preserve the phase characteristics of the input data. As an alternative, Nth-band all-pass based recursive filter with an IIR response is considered as a second candidate. It was observed that using an ALP IIR based DFT modulated filter bank saves in the region of 50\% power in the Coarse Channeliser and 23\% in the overall end-to-end channeliser compared to the Linear-Phase FIR filter based design counterpart. The increased computational cost due to the introduction of reconfigurability and flexibility in communication processors can be reduced by replacing FIR filters with their IIR counterpart implementations without affecting the overall channeliser performance, which was reported in the last section of this paper where the MSE results were very comparable. Although the study was conducted for a particular commercial ADC, It is also possible to use more recent ADCs capable of processing even wider bands. It should be noted this work and the equations presented herewith can easily be scaled to accommodate any other wideband ADC.

\end{document}